\documentstyle[12pt,epsf]{article}
\font\tenbf=cmbx10
\font\tenrm=cmr10
\font\tenit=cmti10

\renewcommand{\large}{\normalsize}

\input{psfig.sty}
\begin{document}

\font\fortssbx=cmssbx10 scaled \magstep2
\hbox to \hsize{
\hfill$\vcenter{\hbox{\bf MAD-PH-1014}
 \hbox{September 1997}}$ }

\vspace{.1in}

\begin{center}
{\tenbf THE HIGHEST ENERGY COSMIC RAYS AND NEW PARTICLE PHYSICS}
\\
\vskip 0.7cm
{\tenrm G.~Burdman and F.~Halzen}
\\[.1cm]
\tenit Department of Physics, University of Wisconsin, Madison, WI 53706, USA
\\[.5cm]
{\tenrm R.~Gandhi}\\
{\tenit Mehta Research Institute, 10 Kasturba Gandhi Marg, Allahabad
211002, India}
\end{center}

\smallskip

{\footnotesize

\begin{center}ABSTRACT \end{center}
{\narrower
It has been argued that the observations of cosmic particles 
with energies in  
excess of $10^8$~TeV represent a puzzle. Its solution requires new
astrophysics or new particle physics. 
We show that the latter
is unlikely given that the scale associated with a new particle
physics threshold must be of order 1~GeV, not TeV and above, in order to
resolve the problem.
In most
cases such new physics should have been revealed by accelerator
experiments. 
We examine the possibility 
that the highest energy
cosmic rays are initiated by non-standard interactions of neutrinos 
in the atmosphere. We show that proposals in this direction either
violate s-wave unitarity or fall short of producing a sizeable effect
by several orders of magnitude.
}}
\section{Introduction}

Cosmic rays reveal Nature's particle accelerators. Ever since the pioneering  
Haverah Park experiment~\cite{Watson} discovered that cosmic particles are
accelerated up to $10^{8}$~TeV energy, the origin of the highest energy cosmic
rays has been hotly debated. Several recent observations of isolated events
with even higher energy is nothing less than paradoxical; they seem to imply  
aspects of particle physics or astrophysics not revealed in previous
experiments. We will outline the puzzle further on.

The energies of such particles exceed by a factor of a hundred million those  
achieved with man-made accelerators. When colliding with atmospheric nuclei,  
the center of mass energy is approximately
500~TeV, more than one order of magnitude larger
than that of the future Large Hadron Collider at CERN. It may therefore seem  
reasonable to speculate that cosmic particles, accelerated to such energy,
may exhibit new particle physics. 
In a  recurrent scenario they are assumed to  
be neutrinos which become strongly
interacting~\cite{domokos,chan} at these extremely high energies.
The physics behind such interactions, being at scales of several tens or even
hundreds of TeV, might be intimately connected to the problem of
flavor and fermion masses.

The main point of our paper is to demonstrate that new GeV, not TeV-scale,
physics is required to have any impact on the problem at hand. This will
follow from the fact that cross sections of several tens of millibarn or
larger must be associated with the new physics. 
It is extremely difficult for such new
thresholds to be turned on up to millibarn cross sections
at  $10^8$~TeV energies without
violating s-wave unitarity. Needless to say that new GeV-scale physics is
unlikely to have escaped the scrutiny of accelerator experiments.

The paper consists of two parts. We first discuss the puzzling features 
of the  
highest energy cosmic rays. 
Subsequently, we study the possibility that 
non-standard 
neutrino interactions at these very high energies 
give an explanation of these events. 
Our focus on neutrinos is motivated by 
the fact that - unlike protons -
they are largely unaffected by attenuation,
as will be discussed in the 
next section. 
We show that, even in the presence of new interactions at high energies,
they cannot provide
a realizable explanation. This will substantiate
our assertion that 
any new particle physics relevant to these issues should have been or can
be revealed in  existing experiments. 
We conclude with some comments.

\section{The Highest Energy Cosmic Rays: A Paradox}

In October 1991, the Fly's Eye cosmic ray detector recorded an event of
energy $3.0\pm^{0.36}_{0.54}\times 10^{8}$~TeV~\cite{flyes}. This event,
together with an event recorded by the Yakutsk air shower array in May
1989~\cite{yakutsk}, of estimated energy $\sim2\times 10^{8}$~TeV, are the
two
highest energy cosmic rays ever seen. More recent papers~\cite{akeno} report  
that the Akeno Giant Air Shower Array, an instrument of over 100
scintillation
detectors spread over a $100$~km$^2$ area, recorded 2 events in the same
energy range.

How Nature accelerates microscopic particles to macroscopic energy is
still a matter of speculation. In order to accelerate a particle to energy
$E$ in a magnetic field $B$, its gyroradius must be contained within the
accelerator. In other words, the accelerator's dimension $R$ has to exceed
the particle's gyroradius $E/B$. This leads to the relation
\begin{equation}
 E \leq BR,
\end{equation}
where the equality can be satisfied for a totally efficient accelerator. It
is generally accepted, that supernovae in our own galaxy accelerate the bulk  
of the cosmic rays, perhaps via shocks driven into the interstellar medium 
by  
the supernova explosions. Although the blueprint of this accelerator is
complex, with a typical size of tens of parsecs and a magnetic field of
several microgauss, its maximum energy reach is easily obtained by
dimensional analysis:
\begin{equation} E_{\rm max} = \left[10^{5}\,{\rm
TeV}\right]\left[B\over3\times10^{-6}\rm\,G\right]
\left[R\over50\,{\rm pc}\right] \,.
\end{equation}
Our own galaxy is too small and its fields too weak to accelerate particles
to energies exceeding $10^{8}$~TeV. This implies that they must be produced
outside our galaxy, possibly near supermassive black holes in active 
galactic  
nuclei where magnetic fields of hundreds of microgauss 
extend over kiloparsec  
distances. The highest energy cosmic rays should point at their sources, 
even  
if they are charged. The gyroradius of a $10^{7}$~TeV proton in the
$3\times10^{-6}$~gauss galactic field is roughly 10~kpc, comparable to the
size of our galaxy. So, $10^{8}$~TeV particles should travel in straight
paths
from their sources through the galactic and intergalactic magnetic fields.

What completes the puzzle is that, at this point, one can reasonably argue 
that the highest  
energy cosmic rays are not nuclei or protons, nor gamma rays or neutrinos,
as long as these particles have standard interactions. 
We present these arguments sequentially:

\vskip 0.4cm
\noindent
$\bullet$
The mean free path of a $3\times 10^{8}$~TeV proton in the cosmic photon
background is only 8.8~Mpc. Protons of this energy, traveling through the
omnipresent 2.7K photon background, will photoproduce pions, and will thus be
demoted in energy over a distance of less than 10~Mpc, i.e.\ much less than
the 100~Mpc plus distance from the posited sources. Alternatively, the
probability for a proton of this energy to traverse 100~Mpc without an
interaction is $1.16\times 10^{-5}$. A cosmic ray proton needs an energy of
$3\times 10^{10}$~TeV to reach Earth from a 100~Mpc source with the observed  
energy.  Needless to say, achieving energies of this order 
becomes a challenge,  
even if the parameters for the standard acceleration mechanisms 
are stretched~\cite{Sigl}.  
From the previous discussion it is clear that the identification of the
highest energy cosmic rays with protons is problematic. The above arguments
apply, {\it mutatis mutandis}, to nuclei.

\vskip 0.4cm
\noindent
$\bullet$
The measured shower profile of the Fly's Eye event is sufficient to conclude  
that the event has not been initiated by a photon. Photons with these
energies
interact in the geomagnetic field, thus starting a cascade well before
entering the atmosphere \cite{McBreen,Aharonian,Vankov}. A Monte Carlo
simulation of the atmospheric shower profile of the Fly's Eye event has been  
performed \cite{vazquez}. The simulation includes interactions with both the  
Earth's magnetic field and nuclei in the atmosphere. They show that a
$10^{8}$~TeV photon encountering the ever increasing geomagnetic field will
interact somewhere between 500 and 10000~km above the Earth's surface. The
most probable height is 3000~km. The dipole magnetic field at this distance
is
roughly 0.1~Gauss. Notice that the shower direction in this
event is almost perpendicular to the field lines. In the primary interaction  
the photon is transformed into a pair of electrons which, subsequently,
suffer
an energy loss as a result of magnetic bremsstrahlung which is peaked 
forward  
at $h\nu/E$~$\sim0.1 $ for $E=10^{8}$~TeV and $H=0.1$ Gauss. The resulting
electromagnetic shower consists, on average, of 6 $\gamma$--rays carrying
65\%
of the primary energy. These $\gamma$--rays of energy $10^{7}$~TeV will
initiate the development of the atmospheric cascade. After further cascading  
the overall photon energy distribution peaks at $10^{5}$~TeV. One must take
into account that at these energies the electromagnetic cascade is elongated  
by the LPM effect \cite{Vankov}.

The bottom line is that a shower initiated by a $3\times 10^{8}$~TeV gamma
ray reaches shower maximum high in the atmosphere at $x_{\rm max}= 1075$ gr
cm$^{-2}$, inconsistent with the observed value of $815 \pm ^{45}_{35}$ gr
cm$^{-2}$. As a result of the large number of secondary photons that
contribute to the composite air shower, the fluctuations are very small. We
conclude that the hypothesis of the event being initiated by a $\gamma$--ray  
is not consistent with the experimental observations. The same conclusion is  
reinforced by the Yakutsk event which is recorded by a giant array of 18
km$^2$. The detector consists of scintillators, \v Cerenkov detectors, muon
detectors and antennas for radio frequency detection. The shower is rich in
muons and therefore not initiated by a $\gamma$--ray.

\vskip 0.4cm
\noindent
$\bullet$
Neutrino origin is also inconsistent with the observed shower profiles. At
these energies the atmosphere is transparent to neutrinos. The ratio of the
neutrino-air and proton-air cross sections is, in the absence of 
new physics,  approximately
$10^6$ at this energy. The particle physics is sufficiently precise to
bracket
its value in the range $10^5 -10^7$. 
This is so even when the energies are so high as to probe very small values
of $x$. 
The average $x$ is given by:
\begin{equation}
<x> = \frac {1}{\sigma} \int^1_0 dx \; x \; \frac{d \sigma}{dx}~.
\end{equation}
It is essential not to neglect the $x$-dependence of the $W$ propagator in the
expression for $d\sigma/dx$ which gives the main contribution to average $x$:
\begin{equation}
\frac{d \sigma}{dx \; dy} = \frac{G_F^2 s}{\pi} 
\left( \frac{M_W^2}{M_W^2+s \; x \; y} \right)^2 x q(x) ~,
\end{equation}
where $s$ is the square of the center of mass energy.
If we assume that  the quark distribution function is given by 
$q(x) \sim 1/x^{1+\epsilon} $, with $\epsilon \sim 1/2$ from
perturbative QCD, 
we obtain 
\begin{equation}
<x> \approx {\cal O}(1)\times
\frac{1}{\sigma}\frac{G_F^2\;M_W^2}{16\pi}\sqrt(M_W^2/s) ~,
\label{avgx}
\end{equation}
with average $Q^2$ of the order $M_W^2$. 
Thus, for $E_\nu\approx (2-3)\times 10^8$~TeV one expects
$<x> \approx (10^{-7}-10^{-8})$. 
However, the fact that this values of $x$ are well below the 
currently measured range, does not represent an obstacle to bound
the neutrino cross section. 
For instance, in Reference \cite{quigg} various methods of extrapolation
at low $x$ are used in order to establish a range for the neutrino
cross section. 
At these energies the charged current cross section varies from 
approximately $ 2\times 10^{-5}$ to $3 \times
10^{-4}$ mb for different structure functions. These are still
very small values.

With a cross section reduced by at least a factor $10^5$ compared 
to protons,  
neutrinos should interact in the earth, not the atmosphere, with relatively  
flat distributions. Although nothing can be made of an odd single event
interacting in the atmosphere, the neutrino scenario is inconsistent with 5
events, or more depending on how one counts, all interacting at the top of
the atmosphere.

We conclude that the highest energy cosmic rays are neither protons or
photons, nor neutrinos. While the data itself rules out photons, 
both protons  
and neutrinos are disfavored by a problematic factor of $10^5$ which
represents the probability that a proton reaches us without attenuation from  
100~Mpc source, and the ratio of the neutrino to proton interaction cross
sections in the case of neutrinos. This is the paradox. Its resolution may
involve new astrophysics, or new particle physics at energies which exceed
those of existing accelerators by two orders of magnitude. 
In what follows we will argue  
that the second possibility is unlikely if we restrict 
the primary to be a known particle experiencing non-standard interactions.
As mentioned earlier, neutrinos are the most promising candidates 
within this option due to the absence of attenuation effects.

\section{Is New Particle Physics the Solution?}

Going the particle physics road is attractive. What if, for instance,
neutrinos became strongly interacting so as to initiate air showers?
Transforming the energy of $10^8$~TeV to the center of mass, yields 
approximately 450~TeV.  
At such energies physics associated with scales as large as $10-100~$TeV may  
be relevant and even dominant.
As mentioned above, this energy scales
might be associated with new particle physics, 
the generation of flavor and fermion masses, 
dynamical supersymmetry breaking, etc. 
The possibility that these new interactions might cause neutrinos
to become strongly interacting at these energies has been raised
in several opportunities. 
For instance, it 
is the underlying physics behind the neutrino
compositeness proposal of Reference~\cite{domokos}.
More recently, a model of spontaneously broken family symmetry 
\cite{chan}, with a typical scale of hundreds of TeV 
and designed to generate flavor, 
was suggested  as a possible origin of 
a very large neutrino coupling at high energies, thus offering 
a potential explanation for  the Ultra High Energy Cosmic
Ray (UHECR) events.
We will now show that these proposals
fail, dramatically.
In order to resolve the puzzle of the highest energy
cosmic rays the new physics scale cannot exceed several GeV. 
On the one hand, 
s-channel unitarity  prevents us from turning on suddenly, at $10^8$~TeV, a
threshold associated with a cross section characterized by a
a typical scale of about $1$~GeV.
More sophisticated proposals might get around the unitarity bound at the 
cost of giving  a very small effect.
We will study below various
specific examples covering these possibilities.

The proton-proton cross section at $10^8$~TeV energy is roughly
100~mb~\cite{bloch}. The interaction length of a proton in the atmosphere
corresponding to this interaction cross section is $40 {\rm ~g\,cm}^{-2}$,
i.e. the
full atmosphere represents 20 interaction lengths. As the interaction length  
is inversely proportional to the cross section, the atmosphere is only 2
interaction lengths for a particle with a cross section of 10~mb. So, in
order
for five cosmic rays to initiate showers near the top of the atmosphere,
their interaction cross section must be several times 10~mb, or not much
smaller than the 100~mb value for protons.

The new particle physics scenarios we consider here are chosen 
partly because  
of the attention each of them has attracted in relation to the UHECR
question. They also span a wide range of models making our conclusions
quite general. Our aim is to show that, with very few and constrained
exceptions, extensions of the standard model of electroweak interactions
at scales above a few TeV cannot be the physics behind UHECR and that the
energy scale necessary to explain the highest energy cosmic rays is
not far above $1~$GeV in most cases. To illustrate this point we will study  
three different classes of models: s-channel resonances, composite neutrinos  
and the t-channel exchange of a gauge boson strongly coupled at high
energies.

We first study the effects of an s-channel $\nu q$ scalar resonance $S$ in  
the $\nu N$ cross section. This is very similar to the study of the effects 
of  
leptoquarks
in UHECR in Reference~\cite{robinett}. The production cross section,
in the narrow width approximation, is given by
\begin{equation}
\sigma(\nu N\to S X)=\frac{\lambda^2\pi}{4M_S^2}
\; x\;q(x=\frac{M_S^2}{s}, Q^2=M_S^2)~,
\label{scalar}
\end{equation}
where $\lambda$ is the coupling of $S$ to quarks and leptons.
In Figure~1 we plot this cross section as a function of the
neutrino energy, for various values of $M_S$ and for $\lambda=1$
\footnote{Normally, leptoquark scenarios have $\lambda\ll 1$}.
For reference, we plot the
SM $\nu N$ charged current cross section, computed using the 
CTEQ4D set of parton distribution functions~\cite{cteq}. 
These are extrapolated 
down to vales of $x$ as low as $10^{-8}$ by using the double logarithmic
approximation~\cite{lowx}. 
The uncertainties associated with the use of this procedure are irrelevant
for the purpose of the calculation of the neutrino cross sections due
to new physics effects, since we are interested in enhancements
of several orders of magnitude.
Also plotted in Figure~1 is the $pp$ cross section, 
which sets the scale a model
must match in order for the neutrinos to interact in the
atmosphere. We observe that in order to obtain a neutrino cross
section of this size at the highest energies the mass scale of the
exchanged particle has to be ${\cal O}(1)~$GeV. 
Of course, such a mass is in
flagrant conflict with all low energy data. 
The idea behind this  simple exercise is to 
show the difficulty of generating a $\simeq 100~$mb cross section
at $E_{\nu}\simeq 10^{12}~$GeV. New particle physics scenarios which  
extrapolate from and extend on established particle physics, cannot generate
neutrino cross sections far above their SM values. In what follows, we
will arrive at the same conclusion in two completely different and seemingly  
promising type of models.

We next consider the possibility that neutrinos are composite with a scale  
$\Lambda_c$
somewhere between $10$~TeV
and several hundred TeV. If the neutrino constituents are colored, they will  
experience strong interactions with quarks and gluons above the
scale $\Lambda_c$. This is essentially the scenario proposed
in \cite{domokos}, where it was suggested that the cross section is
determined by the scale of the strong interactions, $\Lambda_{\rm QCD}$,
as opposed to the scale of compositeness. This would lead to a large
cross section of the order of several millibarns, 
and perhaps to an explanation of
the UHECR events. We will show that this is not the case.
We first notice that the size of the neutrino must be determined
by $\Lambda_c$ and that no color can leak out of a 
$\sim 1/\Lambda_c$ radius. In  
order to resolve the constituents, the wavelength of an exchanged particle  
must be sufficiently small.
In $\nu q$ scattering, this implies that the exchanged gluon can only interact
with the neutrino constituents if its momentum transfer is of the
order of $\Lambda_c$, or larger.
To estimate the neutrino cross section we assume that the
preons inside the neutrino have ${\cal O}(1)$ momentum fractions.
Thus the $\nu N$ cross section is approximately given by
\begin{equation}
\frac{d\sigma}{dx\,dy} \simeq  2\pi\alpha_s \; \frac{s}{Q^4}
\; \left[1+(1+y)^2\right]xq(x)~~,
\label{nucom}
\end{equation}
for momentum transfers satisfying $Q^2>\Lambda_c^2$. In Figure~2 we plot
the neutrino cross section for several values of $\Lambda_c$.
For any reasonable values of  
$\Lambda_c$ the cross section  is nowhere near the $\simeq 100$~mb landmark  
it should reach at $E_{\nu}\simeq 10^{12}$~GeV. The plot of the cross 
section  for
$\Lambda_c=1$~GeV illustrates the fact that this is the relevant
energy scale to enter the millibarn regime, as one would expect. Of course,  
the neutrino compositeness scale is bound by
experiments to be at least a few TeV~\cite{nucomposite}.
The failure of the argument in \cite{domokos} can be traced back to the fact
that color is confined in $r_{\nu}\simeq 1/\Lambda_c$, and therefore the
factor of $Q^4$ in the denominator in  (\ref{nucom}) represents an 
unsurmountable suppression. This feature of  s-wave unitarity prevents
the sudden appearance of a very large effect. 
The statement that the interaction scale should be of about $1$~GeV
is very general and 
can be applied to models where
exotic particles are chosen to be the primary sources of UHECR. 
These must carry color in order to hadronize and thus have a large 
cross section in the atmosphere, regardless of their mass or other
quantum numbers.

Finally, we consider the very intriguing scenario of
Reference~\cite{chan}, where fermions transform under a
spontaneously broken generation symmetry taken to be $SU(3)$.
The generation group is assumed to be dual to $SU(3)$ color.
The massive gauge bosons in this model couple to generation
number with a coupling $\tilde g$, satisfying the duality condition
\begin{equation}
\tilde g g =4\pi~~.
\label{dual}
\end{equation}
These gauge bosons, dubbed ``dual gluons'', induce flavor changing
neutral currents (FCNC) at tree level.
Experimental bounds on FCNC processes force their mass scale to be
at or above the $100~$TeV range. It was pointed out in \cite{chan}
that neutrino interactions could become strong at very high energies
via the exchange of dual gluons, which become strongly coupled
due to the condition (\ref{dual}). This fact explains why there 
would be 
no large effects induced at low energies. 
The $\nu N$ cross section induced by the exchange of a dual-gluon
is given by 
\begin{equation}
\frac{d\sigma}{dx\, dy}= \frac{\pi \;F}{2\alpha_s(Q^2)} 
\; \frac{s}{(Q^2+M_{D}^2)^2} \; xq(x)\left\{ 1+(1+y)^2\right\}~,
\label{csdg}
\end{equation}
where $M_D$ is the mass of the dual gluon and $F$ is a factor of 
order one coming from the group structure of the generation symmetry.
For instance, for $SU(3)$, we have $F=2$ as long as we consider only first
generation fermions in the initial state. 
The $\nu N$ cross section
mediated by dual gluon exchange is plotted in Figure~3 
for several values of the dual gluon mass. It is apparent that for
the desired mass range of $100~$TeV the effect on the cross section is
negligible, even when compared to the SM $\nu$ cross sections.
This is the case despite the very large enhancement coming from the 
running of $\alpha_s$ in the denominator, a consequence of (\ref{dual}).
The main reason for the relative suppression is the value of $M_W/M_D$. 
This is somewhat upset by the fact that the dual-gluon cross section
rises linearly with $E_\nu$ up to very large energies before saturating. 
Even with this feature, the cross section at $E_\nu\approx 10^{12}$~GeV
is about one hundred times smaller that the SM one. 
We see that a dual gluon mass of $50$~GeV, in obvious conflict with
experimental bounds on FCNC, is required in order to yield a sufficiently  
large cross section at the highest neutrino energies.
This mechanism
avoids the need for a ${\cal O}(1)~$GeV scale, given the extreme
strength of $\tilde \alpha(Q^2)$ at very high energies. Even with this  
coupling the model produces an insignificant enhancement of the SM neutrino  
cross section because of the scale of $100~$TeV. On the other hand, \
one could  
in principle imagine a completely unrelated model where the dual 
gluon has no  
FCNC interactions and then is allowed
to be lighter. However, the induced contact interactions, even when
flavor diagonal, are constrained to be governed by a scale above
a few TeV \cite{nucomposite}. Although at these mass scales the
effect of dual-gluon exchange is large compared to the  SM $\nu$
cross sections, it is still several orders of magnitude smaller than
needed to explain the UHECR excess.

We  conclude that it is highly unlikely that neutrino initiated
air showers involving new neutrino interactions are responsible for 
the apparent excess of events in UHECR. We have shown that the needed scale 
is,
in most cases, of ${\cal O}(1)$~GeV which is not an allowed energy scale 
for new neutrino interactions. One type of models that gets around this 
general constraint, does so by having an increasingly strong coupling at
high energies. Even in these cases, the scales that are still allowed 
by low energy constraints (e.g. a few TeV in 
Fig.~2) are already too high to provide a large enough effect.

\section{Some Final Remarks}

We have studied the possibility that the UHECR excess is initiated by
known particles with non-standard interactions at very high energy.
We concentrated on neutrinos as they do not suffer from the 
attenuation that forces protons,for instance , to come from local sources.
We found that, even in the presence of important new physics effects 
at the high energies at hand, neutrino initiated air showers are not 
viable. We have also shown that the energy scale associated with the
interactions responsible for the UHECR should be, in most cases, 
in the vicinity of $1$~GeV. Thus, models postulating exotic primaries 
must arrange for them to form hadrons, which in turn can interact with the
desired cross sections in the atmosphere. An exception to this is the model
of Reference~\cite{domokos}, where the energy scale needed is of the 
order of $100$~GeV due to the large enhancement given by the strength of the 
coupling at high energies.
However, in this as well as in all other cases, the necessary energy scales
are well below the limits allowed by observation. 
We conclude with a few comments about possible alternative 
explanations.

As it can be read 
from Figure~1, leptoquarks \cite{robinett} as well as 
typical supersymmetric models, which are associated
with TeV-scale physics, are irrelevant to cosmic ray issues. At $10^8$~TeV
supersymmetric particles interact with universal electroweak cross section,
i.e. cross sections similar to those of Standard Model neutrinos
~\cite{drees}.

The scenario where the highest energy cosmic rays are light gluinos does not  
violate our no-go argument \cite{farrar}. 
Their mass is indeed in the
GeV-range. But most importantly, they form various supersymmetric
hadrons which interact with the atmosphere with cross sections
governed by the $1$~GeV scale. 
This scenario can be tested by existing accelerator 
experiments~\cite{lgexp}.

Topological defects~\cite{topdef}
are an example of new particle physics not covered by our  
exclusion argument because they are, essentially, a new astrophysical 
source and do not represent new particle dynamics.

Scenarios involving 
exotic primaries, possibly avoiding our arguments, require 
yet additional assumptions in order
to be relevant. While large cross sections with hadrons are  
required, those with photons must be suppressed in order to avoid 
significant  
attenuation in the cosmic microwave background. If not, the new particle has  
properties similar to protons and can only come from local sources.
Heavy stable colored particles fall in this category~\cite{mohapatra}. 
On the other hand, heavy {\em quasi}-stable particles~\cite{berezinsky}
decaying locally, for instance in the halo, 
are not affected by attenuation.

In sum, a particle physics explanation of the UHECR is not viable unless
new interactions {\em and} new matter with the right properties are invoked.
On the other hand,
it is possible that the cosmic ray paradox may have an alternative 
solution which can hardly be
raised to the level of new astrophysics. There may be mechanisms by which
$10^8$~TeV energy is reached locally, not in sources beyond 100~Mpc. Such
speculations have been disfavored. We mention them for completeness: 
galactic  
winds exceeding the size of our galaxy~\cite{jokipii} possibly reaching out  
into the local cluster, and pinball enhancement  of the particle energy
between several galactic supernovae~\cite{axford}.

\section*{Acknowledgments}

The authors thank Chris Quigg for useful comments and discussions.
This research was supported in part by the U.S.~Department of Energy under
Contract No.~DE-AC02-76ER00881 and in part by the University of Wisconsin
Research Committee with funds granted by the Wisconsin Alumni Research
Foundation.


\newpage
\vskip 2cm
\begin{center}
\large\bf Figure Captions
\end{center}

\noindent
{\bf Figure 1:} Neutrino cross section as a function of the neutrino
energy, for the case of scalar s-channel
exchange. For comparison the standard model charged current
neutrino-nucleon cross section, as well as the total pp cross section,
are shown in dashed lines. 
\vskip 0.5cm

\noindent
{\bf Figure 2:}  Neutrino cross section as a function of the neutrino
energy, for the case of neutrino compositeness. 
\vskip 0.5cm

\noindent
{\bf Figure 3:} Neutrino cross section as a function of the neutrino
energy, in the dual gluon model of Reference~\cite{chan}.


\newpage
\begin{figure}[p]
\center
\hspace*{-1.20cm}
\psfig{figure=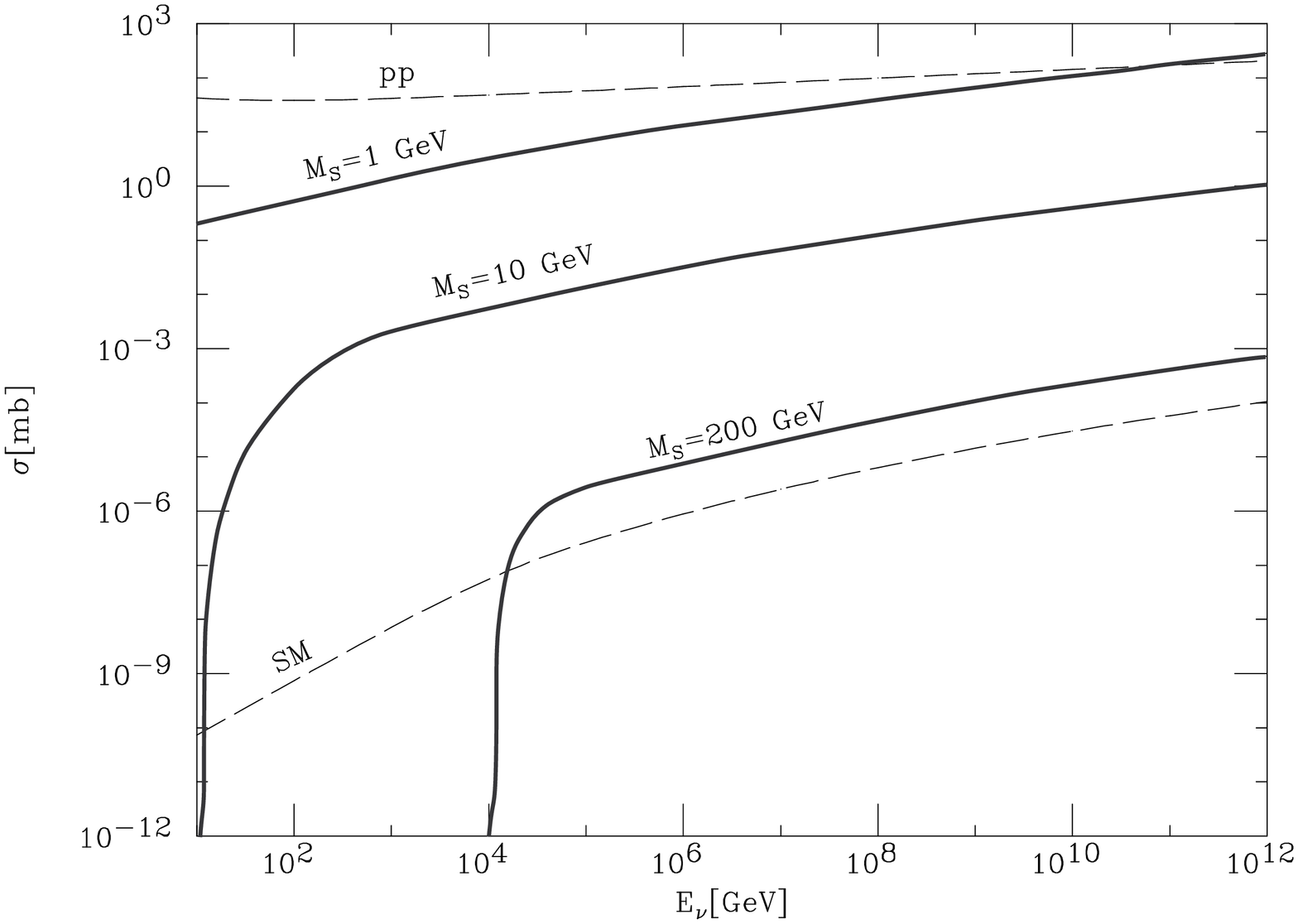,angle=90,height=4.6in}
\vspace*{2.0cm}
\begin{center}
~~~Figure 1
\end{center}
\end{figure}


\newpage
\begin{figure}[t]
\center
\hspace*{-1.20cm}
\psfig{figure=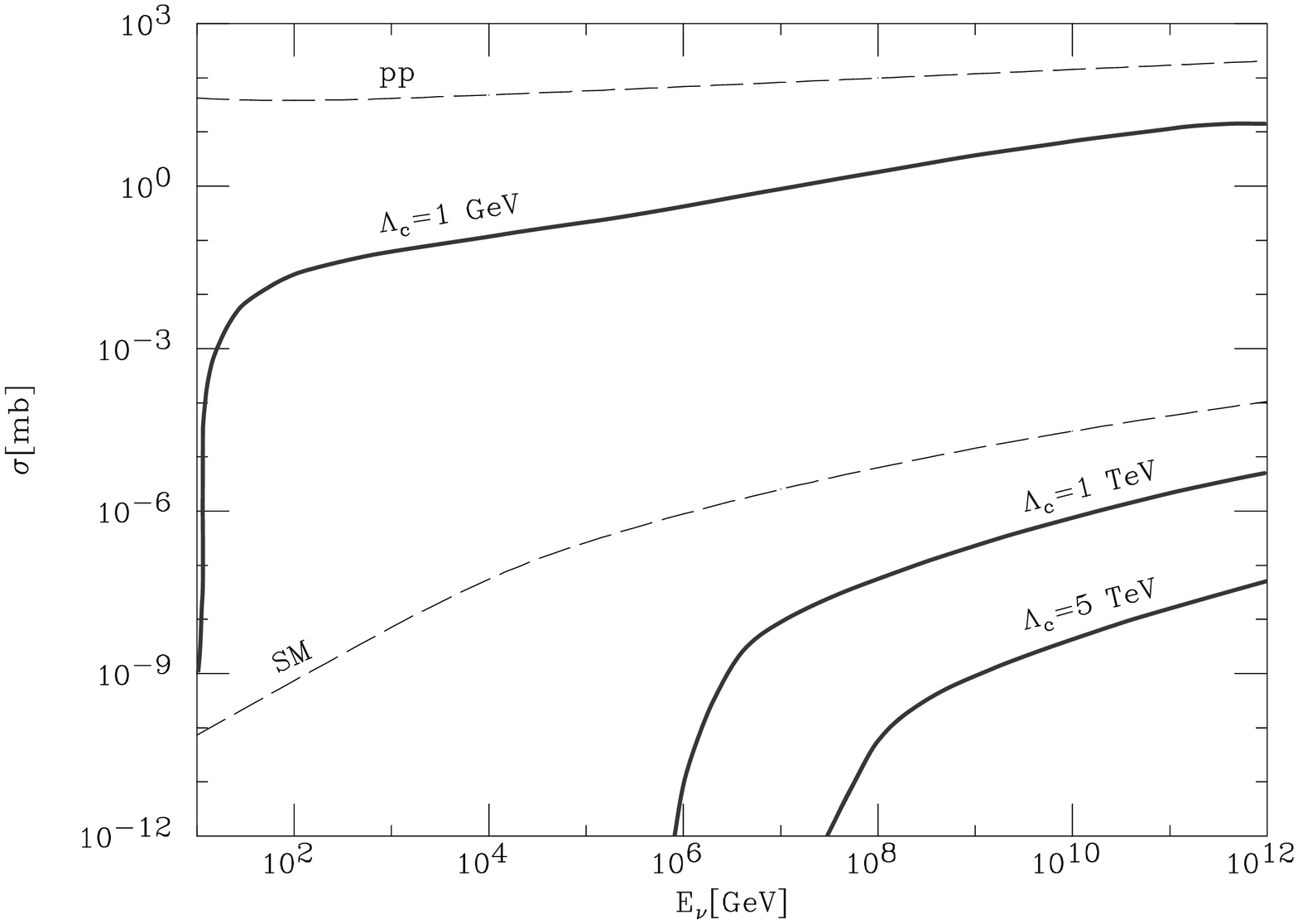,angle=90,height=4.6in}
\vspace*{2cm}
\begin{center}
~~~Figure 2
\end{center}
\end{figure}

\pagebreak
\begin{figure}[p]
\center
\hspace*{-1.20cm}
\psfig{figure=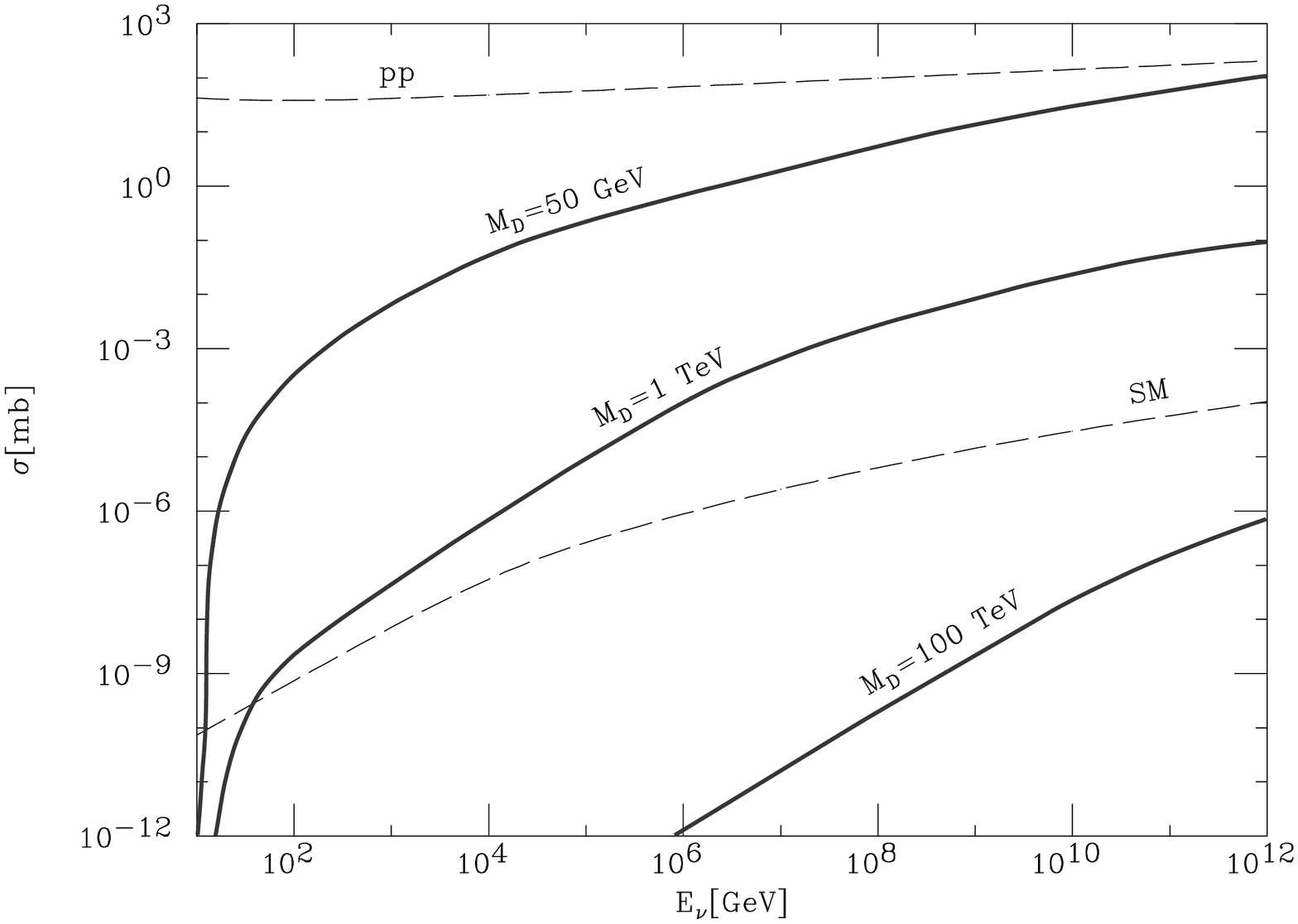,angle=90,height=4.6in}
\vspace*{2cm}
\begin{center}
~~~Figure 3
\end{center}
\end{figure}

\end{document}